{Software}

**Telescope: an interactive tool for managing large scale analysis from mobile devices**

Jaqueline J. Brito[1,†]*, Thiago Mosqueiro[2,†], Jeremy Rotman[3], Victor Xue[3], Douglas J. Chapski[4], Juan De la Hoz[5], Paulo Matias[6], Lana S. Martin[1], Alex Zelikovsky[7,8], Matteo Pellegrini[2], Serghei Mangul[1]*

[1] Department of Clinical Pharmacy, School of Pharmacy, University of Southern California 1985 Zonal Avenue Los Angeles, CA 90089-9121

[2] Institute for Quantitative and Computational Biosciences, University of California Los Angeles, 611 Charles E. Young Drive East, Los Angeles, CA, 90095, USA

[3] Department of Computer Science, University of California, Los Angeles, 404 Westwood Plaza, Los Angeles, CA 90095

[4] Department of Anesthesiology, David Geffen School of Medicine at UCLA, 650 Charles E. Young Drive, Los Angeles, CA, 90095, USA

[5] Center for Neurobehavioral Genetics, University of California Los Angeles, 695 Charles E Young Dr S, Los Angeles, CA, 90095, USA

[6] Department of Computer Science, Federal University of São Carlos, km 325 Rod. Washington Luis, São Carlos, SP 13565-905, Brazil

[7] Department of Computer Science, Georgia State University 1 Park Place, Atlanta, GA, 30303

[8] The Laboratory of Bioinformatics, I.M. Sechenov First Moscow State Medical University, Moscow, 119991, Russia

[†]These authors contributed equally to this work.

*Correspondence: brito.jaque@outlook.com; mangul@usc.edu

**Abstract**

In today's world of big data, computational analysis has become a key driver of biomedical research. High-performance computational facilities are capable of processing considerable volumes of data, yet often lack an easy-to-use interface to guide the user in supervising and adjusting bioinformatics analysis via a tablet or smartphone. Telescope is a novel tool that interfaces with high-performance computational clusters to deliver an intuitive user interface for controlling and monitoring bioinformatics analyses in real-time. Telescope provides a user-friendly method for integrating computational analyses with experimental biomedical research. Telescope is freely available at https://github.com/Mangul-Lab-USC/telescope.

**Introduction**

Exponential growth in the volume of available omics data has reshaped the landscape of contemporary biology, creating demand for a continuous feedback loop that seamlessly integrates experimental biology and bioinformatics[1,2]. Life science and biomedical researchers must choose from an unprecedented diversity of software tools and datasets designed for analyzing increasingly large outputs from modern genomics and sequencing technologies, which are supported by high-performance cluster infrastructures[3]. Scientific discovery in academia and industry now relies on the seamless integration of bioinformatics tools, omics datasets, and large clusters[4–8].

Many life science and biomedical researchers lacking computational training now must learn how to use computational tools in order to process data from their experiments or seek broad patterns in omics data. Ideally, any bioinformatics analysis tool should provide an easy-to-use interface through which the researcher can run and monitor each analysis of omics data[9]. A friendly user interface for omics tools would also enable the researcher with limited computational background to monitor and adjust their analysis without intervention. Lack of user interface management tools pose an obstacle to novice users to perform analysis on of high-performance computing clusters[10]. The procedure of connecting to the cluster often involves a multi-step process and requires generating SSH keys or other forms of authentication. The necessity of using the UNIX command line for each step may discourage potential users.

Yet most bioinformatics tools require the researcher to spend a large amount of time manually adjusting and supervising actively running analytical tasks (referred to as jobs) via command line in a computational pipeline. Today's high-performance computational facilities are capable of processing considerable volumes of data, but a new bottleneck has developed: their user interfaces require of the researcher fluent knowledge of the command line in order to manipulate analysis in real time.

There is a pressing need to seamlessly integrate bioinformatics analysis into the experimental analysis performed by biomedical research, in order to expand research opportunities to individuals lacking a computational background and to reduce the time burden of any researcher who uses a computational pipeline. One example in this direction is the Galaxy Project, which provides a friendly and interactive interface to deploy simple bioinformatics pipelines[11]. Despite many advantages, Galaxy Project lacks a flexible interface to manage the analytical tasks and many parameters related to allocating the computational resources predified (i.e., the number of processes is hard coded)[12].

Bridging the gap between bioinformatics and biological experimentation requires an on-the-fly job management application that is fronted by a user-friendly interface[13]. We developed Telescope to addresses this challenge. Telescope is capable of leveraging common and familiar technologies that do provide a user-friendly interface to manage jobs from any mobile device without compromising flexibility for advanced users. For example, Telescope allows users to track with their smartphones any bioinformatics tools (e.g., GATK[14]) or jobs submitted by specific platforms (e.g., Galaxy Project[14,15]), displaying in real-time the partial outputs, warnings, and error messages associated with each job. Telescope includes the following functionality:

- tracking the progress and performance of actively running bioinformatics tools;
- displaying in real-time the current output of an active job;
- interacting with the computational cluster with minimal effort, allowing cancellation and/or rescheduling of jobs with different parameters, or new job queuing;
- using statistics archived from previous jobs to estimate the resources necessary for future jobs.

Telescope is designed to natively operate with a simple and straightforward interface based on Web 2.0 technology that is compatible with most modern devices (e.g., tablets and smartphones). Moreover, Telescope assumes little from the server side: the existence of a scheduling system (e.g., Sun Grid Engine, SLURM[16]) and SSH connection, both elements featured in virtually all cluster systems dedicated to high-performance computing. As no further assumptions are made, Telescope is tuned to interfere as minimally as possible with cluster performance. We successfully tested Telescope at UCLA's campus-wide computational cluster[17], and we designed the tool for smooth integration with other cluster systems. In order to integrate Telescope in high performance clusters, the technical team managing the cluster must only review Telescope's requirements.

**Related Work**

Several tools exist that provide management and monitoring of bioinformatics analysis tasks, but they offer limited functionality and deployment when compared to Telescope. PHPQstat[18] and GE Web Application[19] are open-source PHP applications that provide web interfaces which allow users to monitor the status of jobs managed by Sun Grid Engine (SGE), a commonly used high throughput cluster system. PHPQstat and GE Web Application are limited to use with SGE and display only details of the jobs currently running on the cluster. (Telescope includes in the display for each job additional functionalities, such as job submission and tracking history.) Virtual Desktop (VDI)[20] provides users a web-based user interface to interact with the FASRC Cluster at Harvard University. Among other functionalities, VDI allows users to check the status of a job, edit an existing job, and submit new jobs. However, VDI is proprietary software that is limited to deployment on the FASRC Cluster; implementation details are not publically available.

Applications of distributed processing frameworks, such as Apache Spark[21] and Hadoop MapReduce[22], can be monitored via the framework's web-based user interfaces. These tools display detailed information about each job, including the worker nodes, statuses of job stages, and memory usage. Applications such as Apache Spark, Hadoop, and MapReduce are specifically designed for each framework and are incapable of working with commonly used scheduling systems like SGE or individual cluster systems managed by universities.

Several existing tools can be used to create and monitor jobs using a web-based interface, but support only specific programming languages or processing pipeline formats. For example, Luigi[23] is a Python module that can be used to manage jobs via the internet. Airflow[24] allows the creation of DAGs (Directed Acyclic Graphs) that specify a pipeline for processing of tasks; it also provides a user interface that allows users to visualize the processing status of the jobs specified by the DAGs. Compared to these tools, Telescope is a more general tool because its main objective is to leverage the common existence of scheduling systems (e.g., SGE) on clusters. Thus, Telescope is not designed for nor is restricted to a specific programming language or processing pipeline format. Telescope was initially developed to work with SGE, but it is designed to be configurable to other scheduling systems.

Finally, several tools enable an interactive approach to building and executing bioinformatics analysis tasks, but lack a function that allows the user to remotely monitor jobs. Jupyter Notebooks, an open-source web application that supports the creation and sharing of live code and data visualizations, allow users to connect to clusters and run jobs using web browsers[25,26]. However, the Jupyter Notebooks system does not allow the user to monitor jobs from a mobile device.

**Methods**

Telescope is comprised of two main features (Figure 1): a mobile-friendly user interface that relies on Web 2.0 and a connection to SSH-enabled servers. Telescope gathers job information through a Job Manager which connects to the target cluster via the Connection Manager. Job information is then stored in Telescope's Local Database to support job analytics and a searchable history. The User Interface relies primarily on both the Local Database and the Rate Limiter to render all relevant job information into a mobile-friendly web page while limiting the impact of Telescope's interaction with the target cluster. In the following sections, we describe Telescope's key components in detail.

**Job Manager.** This component handles all job requests. The Job Manager supports the operation of checking a job's status, cancelling an existing job, and creating a new job. Given a cluster's specific scheduler manager, the Job Manager leverages automated code generation based on the input data. The generated code is routed to the Connection Manager, which leverages SSH's secure code execution capability. The Telescope Core then stores the results from a completed job in the Local Database.

**Connection Manager.** This component interfaces with the target cluster. The Connection Manager establishes communication via an SSH connection using key pairs for authentication. Telescope then leverages this connection to exchange discrete messages with the cluster server. As the messages are encrypted using the industry-standard SSH protocol, Telescope is able to gather information without compromising the user's privacy. The Connection Manager also stores any SSH keys provided by the user.

**Local Database**. The Local Database keeps records of all monitored jobs. An entry is created for each job and archives the job id, job name, and user login. The Local Database also stores information regarding the requested resources (e.g., number of cores requested, memory requested), the current status of the job, and relevant metrics (e.g., elapsed time, max peak memory). The stored attributes can be configured for different scheduling systems (Table S1 lists the attributes in the table Job assuming a cluster with SGE). These records are retained over time to support job statistics and analytics. As this data is aggregated, the average memory and elapsed time for a given bioinformatics pipeline may be extracted as a function of the input parameters.

**Status Scheduler.** For each job monitored by Telescope, the Status Scheduler periodically checks the cluster to update the Local Database with the most recent status data. The Status Scheduler is a background process and triggers update requests for all jobs in predetermined time intervals. These updates are performed in two steps. First, Telescope issues a query to obtain a list of all jobs running in the cluster. Then, for each active job, a new query requests detailed information. For *ad hoc* requests from a user, only this user's jobs are inspected.

**Telescope Core**. The Telescope Core interconnects all components in the Telescope application. The User Interface and Status Scheduler generate job requests that are sent to the Telescope Core, which the Job Manager receives and handles. The results of completed job requests are propagated to update the Local Database, User Interface, and Cache. Telescope employs a Rate Limiter to restrict the rate of requests running under a specified threshold, which prevents an overload of the system running Telescope and, more importantly, the target computational cluster. (Rate limiting is a common technique used to prevent denial of service (DoS) attacks[27].) Each user request must pass through this limiter before reaching the Telescope Core. When the current rate of job request exceeds the maximum threshold, additional user requests are sent to the Cache, which maintains the results of the user's last requests, rather than the Telescope Core. In addition, Telescope applies an exponential back-off algorithm that increases the time interval during which the system can accept another request from the same user.

**User Interface**. Users interact with Telescope through a mobile-friendly web interface (Figure 2). User authentication when logging into Telescope is performed via the OAuth protocol[28], which conducts verification using the user's existing accounts from popular internet services (e.g., Google, Twitter, Facebook). After logging into Telescope, the initial web page displays a summary of all jobs actively running on the cluster under the user's account (Figure 2, left panel), including the job identification code and name, username, current state, and starting timestamp. Each job ID is linked to a page containing more specific data for that job (Figure 2, right panel), including the name of the script file and directory, the content of the script file, and the last few available lines from the output file. Warnings and error messages are collected from the content of logs, defined on .e files. In addition to visualizing jobs that are queued, users can also cancel or create new jobs via the User Interface. Therefore, the User Interface also supports inputting parameters to pre-defined bioinformatics pipelines.

**Security.** Because Telescope handles private information and SSH keys, we designed a system that leverages industry standards for data handling and mitigation of vulnerabilities. Stored SSH private keys are encrypted using PBKDF cryptography, as recommended by the Public-Key Cryptography Standards (RFC 8018)[29]. In cases where a private key is compromised, Telescope users may initiate a key revocation policy. Telescope currently supports SSH key revocation by deleting the compromised SSH fingerprints and updating the revocation list, a procedure that covers most Linux distributions. If a custom security policy is required by a user or cluster administration team, Telescope's modular implementation can be easily tailored by Telescope administrators.

**Discussion**

Telescope interacts with the computational resources directly, at the operating system level, and spares the user from learning in-depth computer science material or devoting substantial time to manually interacting with the computational pipeline. Telescope is domain agnostic and can be used by anyone performing extensive computational analyses (e.g. deep learning, large-scale simulations for climate research).

Data retained in the database could support analytics and generate insights about job behavior, enabling users to predict resource allocation and forecast computation time. We are working on expanding the prediction feature with a simple, automated mechanism based on regular expressions that allows users to attach tags to jobs that can later be used for aggregations and analytics. For instance, data of previous jobs of read alignment tools (**Figures S1-S2**) (**Supplemental Note 1**) stored in table Job (**Table S1**) could have been tagged with tool name and number of reads. Then, Telescope would be able to estimate the expected elapsed time and max amount of memory required to run these tools as a function of the number of reads. In the future, we are planning to systematically collect information about the computational resources of the jobs run through Telescope. We will use recorded information to develop and provide a template that allows users to choose potential software and processing types to make better choices for resource usage.

Telescope has an intuitive user interface and demands minimal requirements from the computational cluster, making the tool appealing to users lacking a computational background, who often face a steep learning curve to operate computational resources, and to experienced users who often manage a large number of jobs and repetitive tasks. As computational clusters run Unix-based operating systems, Telescope does not eliminate completely the interaction with command line prompts, but contributes to lower the bar needed to effectively run and monitor bioinformatics analyses at scale.

We observed that Telescope users who are new users of Unix operating systems are able to, within seconds, check the status of a job and look for warning and error messages: as fast as opening their web browsers and connecting to Telescope. By addressing the challenges inherent to learning command line, Telescope was designed to invite users with any level of computational experience to join the bioinformatics community.

The development of Telescope demonstrates that the current model where bioinformatics analyses are outsourced with no control during job execution (for example, use of pre-cut pipelines wrapped in Graphical User Interfaces) is inefficient and prevents biomedical investigators from harnessing the true potential of their computational tools in the wet lab environment. While Telescope does not directly improve the runtime performance of bioinformatics tools, the application increases accessibility of biomedical data analyses to the scientific community and provides for all users a tool for improving work productivity. Real-time tracking allows biomedical researchers to access partial results—before the analytical task has been completed on a large dataset—and identify potential problems with the analysis or sequencing experiment.

The ideas and results presented in this study represent a contribution toward mitigating the digital divide in contemporary biology. By offering real-time job management tracking and control over computational clusters even on mobile devices, Telescope can help researchers accomplish a seamless feedback connection between bioinformatics and experimental work with minimal performance interference.

*Declarations*

**Ethics approval and consent to participate**

Not applicable.

**Consent for publication**

Not applicable.

**Availability of data and materials**

The software presented in this paper is freely available at

https://github.com/Mangul-Lab-USC/telescope.

Telescope is registered at bio.tools and SciCrunch.org databases as https://bio.tools/Telescope

and RRID (SCR_017626), respectively.

**Competing interests**

The authors declare that they have no competing interests.

**Funding**

T.M. acknowledges support from a UCLA QCBio Collaboratory Postdoctoral Fellowship and the QCBio Collaboratory community directed by Dr. Matteo Pellegrini. A.Z. has been partially supported by NSF Grants DBI-1564899 and CCF-1619110 and NIH Grant 1R01EB025022-01.

**Authors' contributions**

T.M. proposed and scoped the project. J.J.B. and T.M. developed the software presented in this paper and were major contributors in writing the manuscript. J.R., V.X., D.J.C., J.D.H., P.M.,

L.M., A.Z. and M.P. contributed to portions of the code and in writing the manuscript. S.M. lead the project and contributed in writing the manuscript.

**{Figure Legends}**

**Figure 1.** Telescope Architecture. The Job Manager gathers job information by connecting to the target cluster via its Connection Manager. Telescope's Local Database keeps records of this information, which is rendered by the User Interface into a mobile-friendly web page.

**Figure 2.** Telescope User Interface. The first screen displays the status of the jobs on the cluster. The next screen shows detailed information about the first listed job: source directory, name and content of the script file, and last lines of current task output.

**Figures**

Figure 1.

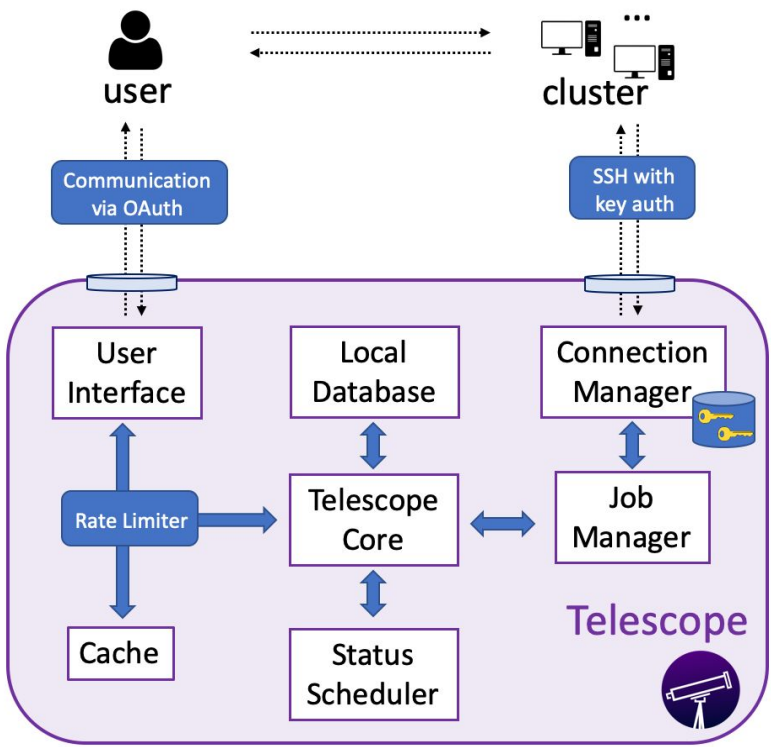

Figure 2.

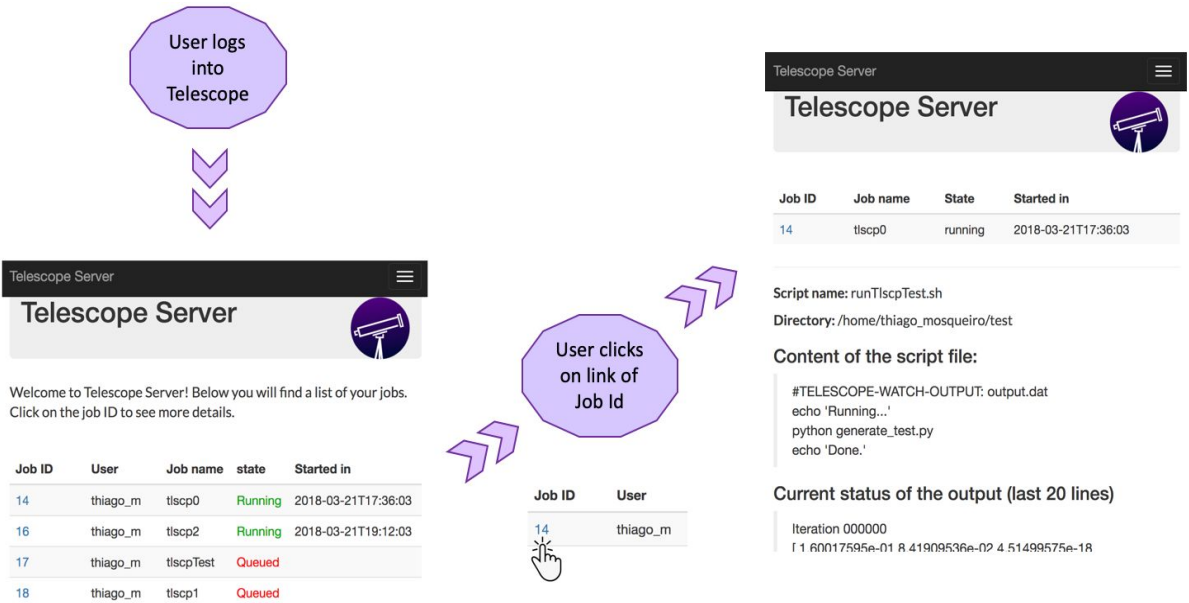

**Supplemental Tables**

**Table S1.** Local Database's schema, explicitly listing attributes, their corresponding data types, and descriptions. These attributes correspond to the information provided by the qstat function of scheduling system Sun Grid Engine.

| Attribute name | Data type | Description |
| --- | --- | --- |
| jobId | INTEGER (PRIMARY KEY) | Unique job id |
| jobName | TEXT | Name of the job |
| user | VARCHAR(30) | Username |
| status | INTEGER | Current status of job (last time qstat was updated) |
| path | TEXT | Path to the script that is being run |
| command | TEXT | Command used to submit job |
| sourceDirectory | TEXT | Directory from which the job was submitted |
| outpath | TEXT | Path and name for output file |
| memoryRequested | TEXT | Amount of memory requested |
| parallel | INTEGER | Running in parallel (1) or not (0) |
| cores | INTEGER | How many cores requested? |
| timeAdded | VARCHAR(30) | When was this entry added to the database? |
| runTime | TEXT | Time job has been running on cluster |
| timeRemaining | TEXT | Time remaining before job is killed by cluster |
| currentMemory | INTEGER | Memory currently in use by job |
| maximumMemory | INTEGER | Maximum memory used so far in job's history |
| clusterNode | TEXT | Node on which job was run |
| finalRunTime | TEXT | For finished jobs, how long did they run? |
| finalStatus | TEXT | How did the job end? Completed, killed, aborted? |

**Supplementary Figures**

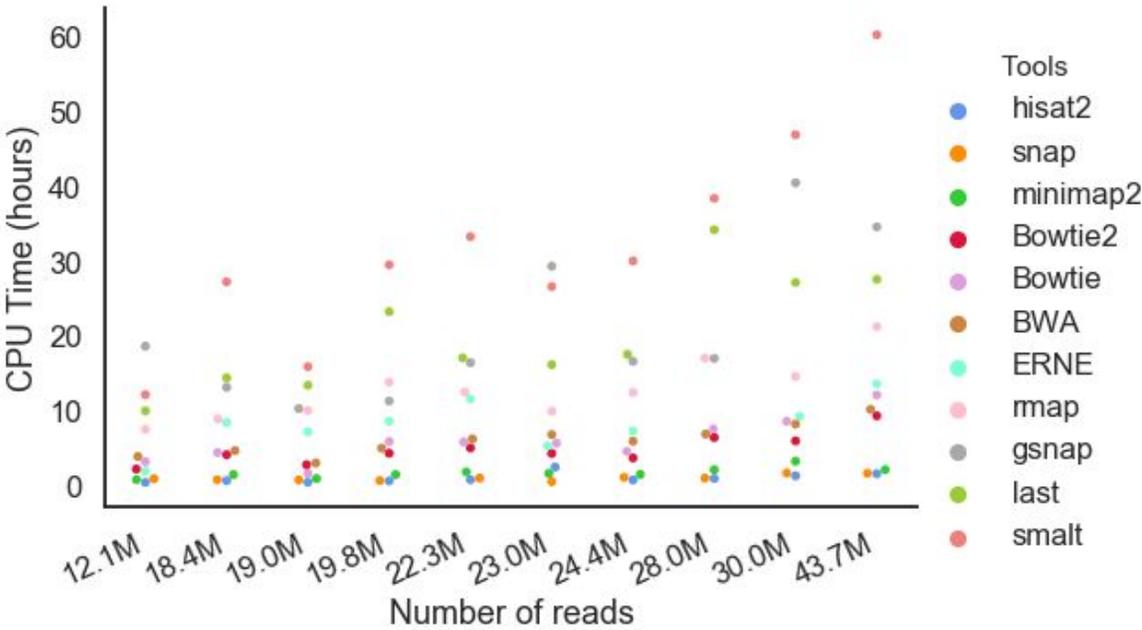

Figure S1. Comparison of the runtime (measured by CPU time in hours) for each tool against the size of each sample (measured by the number of reads).

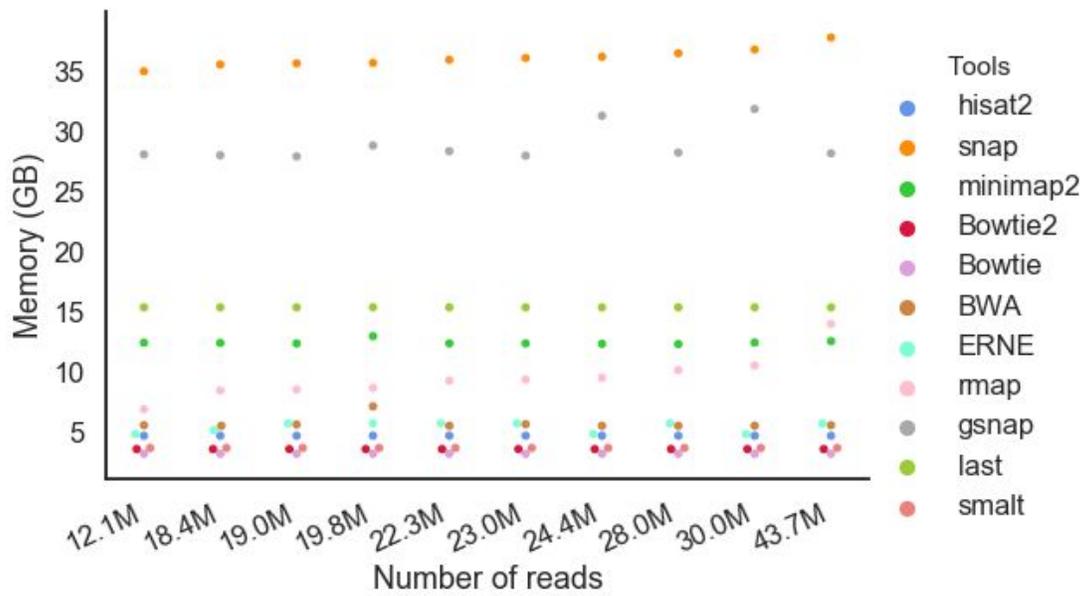

Figure S2. Comparison of the RAM (measured in gigabytes) used by each tool against the size of each sample (measured by the number of reads).

**Supplemental Note 1**

A number of choices can affect a job's impact on cluster resources, including the specific bioinformatics analysis tool used and the size of the input omics dataset. To investigate the effects these specific choices have on cluster resources, we downloaded 11 read alignment tools available through Bioconda: Bowtie, Bowtie2, BWA, ERNE, gsnap, hisat2, last, minimap2, rmap, smalt, and snap. Each tool was used to align 10 whole genome sequencing (WGS) samples from the 1000 Genomes Project. All of these samples are available through the NCBI sequence read archive (SRA) with the following accessions: ERR009309 (12.1M reads), ERR013127 (23.0M reads), ERR013138 (30.0M reads), ERR045708 (43.7M reads), ERR050158 (19.8M reads), ERR162843 (28.0M reads), ERR181410 (22.3M reads), ERR183377 (18.4M reads), SRR061640 (19.0M reads), and SRR360549 (24.4M reads). For each sample and tool combination, we recorded the CPU time (Figure S1) and the RAM (Figure S2) required by the job. Number of reads was calculated by considering two Illumina sequencing paired ends as a single read.